\documentclass[aps,twocolumn,raggedbottom,prl,showpacs,nobalancelastpage,amssymb,groupedaddress]{revtex4}
\usepackage{graphics} 
\usepackage{epsfig}
\begin{document} 
\title{\Large \bf  Quantized dynamics of a coherent capacitor} 
\author{M. Moskalets$^{1,2}$, P. Samuelsson$^{3}$ and M. B\"uttiker$^1$}
\affiliation{
$^1$D\'epartement de Physique Th\'eorique, Universit\'e de Gen\`eve, CH-1211 Gen\`eve 4, Switzerland\\
$^2$Department of Metal and Semic. Physics, NTU "Kharkiv Polytechnic Institute", 61002 Kharkiv, Ukraine\\
$^3$ Department of Physics, University of Lund, Box 118, SE-22100,
Sweden}
\date\today
\begin{abstract}
A quantum coherent capacitor subject to large amplitude pulse cycles can be made to emit or reabsorb an electron in each half cycle.  
Quantized currents with pulse cycles in the GHz range have been demonstrated experimentally.  We develop a non-linear dynamical scattering theory for arbitrary pulses to describe the properties of this very fast single electron source. Using our theory we analyze the accuracy of the current quantization and investigate the noise of such a source. Our results are important for future scientific and possible metrological applications of this source.
\end{abstract}
\maketitle
\begin{figure}[t]
  \vspace{0mm}
  \centerline{
  \epsfxsize 7cm
   \epsffile{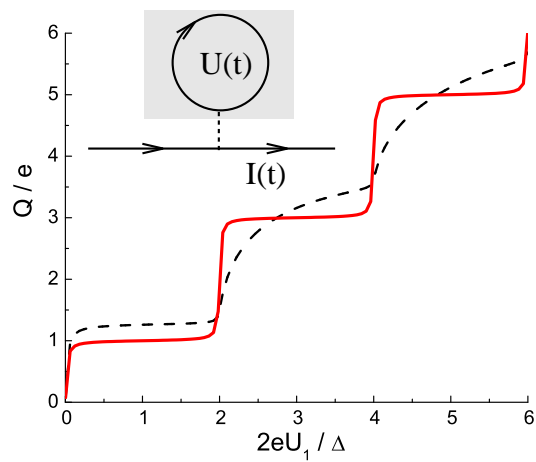}
             }
  \vspace{0mm}
  \nopagebreak
  \caption{ (Color online)
Inset: Quantum capacitor of Ref.\cite{Gabelli06,Feve07}. Electron motion is along an edge state. Transmission through the QPC is shown as a dashed line. The hatched rectangle is a back-gate inducing the potential $U(t)$. Main: The charge $Q$ (red solid line) emitted for half of a period as a function of the (large) amplitude of a slow excitation voltage $U(t) = U_{1}\cos(\Omega t)$. The black dashed line is the charge calculated using only the first harmonic of the current. Parameters are: $k_B\theta = 0$, $T=0.1$.  At $U_1 = 0$ the Fermi energy coincides with one of the energy levels of the dot.
}
\label{fig1}
\end{figure}

{\it Introduction}.-- 
Recent experiments have demonstrated two quantization phenomena in mesoscopic 
capacitors \cite{Gabelli06,Feve07}. The capacitor consists of a quantum dot which can exchange carriers only via one quantum point contact with an electron reservoir. Transport is induced by varying the voltage at a gate to which the dot is coupled capacitively (Fig.\ref{fig1}). 
The first quantization phenomenon is observed in the charge relaxation resistance $R_q$
which together with the capacitance determines the $RC$-time. 
For small amplitude, sinusoidal voltages, the experiment of Gabelli et al.\cite{Gabelli06}  confirmed an earlier prediction \cite{BTP93} and demonstrated a charge relaxation resistance $R_q$ quantized at half a resistance quantum. A necessary condition is a contact which permits transmission of a single (spin polarized) quantum channel. As long as electron motion in the dot is coherent, the quantization of $R_q$ holds for arbitrary values of the transmission probability. This is remarkable. 
Whereas the quantization of the Hall resistance both in the integer \cite{intqhe} and fractional \cite{fracqhe} quantized Hall effect, as well as the quantization of conductance in a ballistic contact \cite{ballistic} are due to perfect chiral transmission channels for which back scattering is suppressed \cite{mb1988}, the quantization of the charge relaxation resistance has an entirely different origin. 
It is due to the fact that the mean square dwell time of a single scattering channel is equal to the square of the mean dwell time \cite{levens}. This equality holds independently of the degree of back scattering in the channel and is valid in the presence of 
interactions \cite{NLB07}. At frequencies larger than the inverse RC-time, the capacitor can respond inductively \cite{indWWG07,indGabeli07}.

In a subsequent experiment F\`{e}ve et al.~\cite{Feve07} revealed a second quantization effect. They subjected the capacitor to square pulses with amplitudes sufficiently large to drive a level of the capacitor through the Fermi energy. In this non-linear regime, for an appropriate pulse voltage, it was found that during the first half cycle an electron is emitted from the capacitor and during the second half cycle of the pulse an electron is reabsorbed. 
As a consequence the first Fourier component of the alternating current is quantized and given by  
\begin{equation}
\label{mb_1}
I_{\omega} = 2\,e\,f 
\end{equation}
with $f = 1/{\cal T}$ where ${\cal T}$ is the duration of the pulse. 
Amazingly the quantized current was observed  for frequencies in the GHz range 
and is thus large (of the order of  several hundred pA). 
Such a fast quantized electron source might be useful in testing systems against the addition or removal of a single electron. It might be useful for metrological purposes \cite{tria,recent} and it might be interesting for quantum computation schemes that use single electrons generated on demand similar to linear optical quantum computation schemes \cite{knill}.

{\it Model}.--
Quantization phenomena, especially in electrical transport \cite{intqhe,fracqhe,ballistic}
play an important role even beyond the purely scientific domain. It is thus desirable to provide a theoretical understanding of the novel quantization phenomena revealed by the recent experiments \cite{Gabelli06,Feve07}.
The quantized current, Eq.(\ref{mb_1}) is obtained for a large amplitude voltage pulse. The theoretical challenge is thus to provide a treatment of the non-linear dynamical response of a quantum coherent capacitor.
Below we present such a treatment for a model \cite{Gabelli06,Feve07,PTB96} consisting of a single circular edge state 
of circumference $L$ coupled via a quantum point contact (QPC) to an edge state which is connected to a reservoir of electrons with temperature $\theta$ and the Fermi energy $\mu$, see inset of Fig.\ref{fig1}.
A periodic in time potential $U(t)=U(t+{\cal T})$ is applied uniformly over the quantum dot.
We determine the non-linear alternating current $I(t)=I(t+{\cal T})$.

We use the Floquet scattering matrix approach of Ref.\cite{MBstrong02} for noninteracting electrons which has proved useful to treat the problem of quantum pumping \cite{pump}. 
The Floquet scattering matrix element $S_{F}(E_{n},E)$ is an amplitude for an electron which enters the conductor with energy $E$ and exits the conductor with $n$ additional modulation quanta  $n\hbar\Omega=E_{n}-E$ with $\Omega=2\pi/{\cal T}$ and where ${\cal T}$ is the period of the time-dependent potential $U$. 
For a time-independent potential, the relevant energy scale of our scattering problem is the level spacing $\Delta = h/\tau$ of the dot, with $\tau=L/v_D$ being the time taken by a carrier to drift with velocity $v_d$ along the full circle of the edge state inside the quantum dot. 
In contrast, the scattering amplitudes $r$ and $\tilde t$ for reflection at the QPC can be taken to be energy independent. 
Therefore, we can distinguish adiabatic and non-adiabatic regimes depending on whether $\Omega\tau\ll 1$ or $\Omega\tau\gg 1$.

{\it Adiabatic current}.--
We first consider the adiabatic regime, $\Omega \tau \ll 1$, where a simple and physically intuitive expression for the current can be found, a detailed derivation is presented below. 
We find that the time-dependent current can be expressed in terms of the instantaneous density of states $\nu(t,E) =\nu_{0}(E-eU(t))$ with $\nu_0(E)= 1/(2\pi i)S_{0}^{\star}(E)\partial S_{0}/\partial E$ being the density of states 
and $S_{0}(E)$ being the scattering matrix of a stationary dot. 
Thus $\nu(t,E)$ is the density of states frozen at time t. 
The current is the sum of a capacitive $I_{c}$ and dissipative  $I_{d}$ contribution simply given by, 
\begin{equation}
\label{mb_2}
\begin{array}{c}
I_{c}(t) = e^2\int dE \left( -\frac{\partial f_0}{\partial E} \right) 
\nu(t,E) \frac{\partial U}{\partial t} ~, \\
\ \\
I_{d}(t) = - e^2\int dE \left( -\frac{\partial f_0}{\partial E} \right)  
\frac{h}{2}\frac{\partial}{\partial t}  \left[\nu^2(t,E) \frac{\partial U}{\partial t} \right]. 
\end{array}
\end{equation} 
Here $f_{0}(E)$ is the Fermi distribution function. Terms of order $[\Omega\tau]^3$ and higher are neglected. 
Importantly, using the general adiabatic expansion for the Floquet scattering matrix \cite{adia}, one can show that Eq.(\ref{mb_2}) remains valid for an arbitrary spectrum of a capacitor.
Eq. (2) is the main result of the paper.

To gain insight into Eq.~(\ref{mb_2}) we introduce a simple electrical circuit to model our system. It is a capacitor and a resistor coupled in series and subject to a voltage $U(t)$. 
Generally such a circuit is a non-linear one.
Therefore, it is convenient to introduce a differential capacitance $C_{\partial} = \partial Q/\partial U_{C}$ (where $Q$ is a charge and $U_{C}$ is the voltage on the capacitor) and a differential resistance $R_{\partial}=\partial U_{R}/\partial I$ (where $U_{R}= U - U_{C}$ is the voltage drop on the resistor). 
Then the low frequency current in this circuit is:
$I(t) = C_{\partial}\partial U/\partial t - R_{\partial}C_{\partial}\partial[C_{\partial}\partial U/\partial t]/\partial t$. 
Comparing it with Eq.~(\ref{mb_2}) we find:
\begin{equation}
\label{mb_3}
\begin{array}{c}
C_{\partial}(t) =  e^2 \int dE \left( -\frac{\partial f_0}{\partial E} \right) 
\nu(t,E) , \\
\ \\
R_{\partial}(t) = \frac{h}{2e^2} \int dE \left( -\frac{\partial f_0}{\partial E} \right) \frac{\partial}{\partial t} \left[\nu^2(t,E)\frac{\partial U}{\partial t} \right] \\
\times \left\{ \int dE \left( -\frac{\partial f_0}{\partial E} \right) \nu(t,E)  \int dE \left( -\frac{\partial f_0}{\partial E} \right) \frac{\partial}{\partial t} \left[\nu(t,E)\frac{\partial U}{\partial t} \right] \right\}^{-1} .
\end{array}
\end{equation}
These time-dependent quantities generalize the linear response capacitance $C_{q}$ and charge relaxation resistance $R_{q}$ of a mesoscopic capacitor introduced in Ref.~\cite{BTP93} and experimentally investigated in Ref.~\cite{Gabelli06}, to the non-linear regime.
Thus, in the non-linear regime, the frozen density of states $\nu(t,E)$ defines a differential capacitance in the same way as the stationary density of states $\nu_{0}(E)$ defines a quantum capacitance: $C_{\partial} (t) = C_{q}(t)$. 
In contrast, the non-linear differential resistance can not be simply related to a linear response one: $R_{\partial}(t) \ne R_{q}(t)$.
However, as will be shown below, the noise generated is proportional to $R_{q}(t)$ rather than $R_{\partial}$.

Eq.~(2) allows us to analyze the general conditions for the appearance of a quantized alternating current. 
This current corresponds to emission (absorption) of an integer number of electrons for half of a period when the driving voltage changes from a minimal $U_{min}$ to a maximal $U_{max}$ value. 
To find the charge $Q$ flowing out of the capacitor we integrate Eq.~(\ref{mb_2}) over half of a period and find to leading order in
$\Omega$:
\begin{equation}
\label{mb_4}
Q = Q_{d}(U_{min}) - Q_{d}(U_{max}) + {\cal O}(\Omega\tau), 
\end{equation}
where $Q_{d}(U) = e\int dE f_0(E) \nu_{0}(E-eU)$ is the charge of the dot in a stationary potential $U$.
Thus for slow driving the emitted charge is directly related to the change of the frozen charge on the dot.
With increasing frequency this relation breaks down.  
If the electron levels in the dot have a small width 
and the temperature is low enough 
then the emitted/absorbed charge is quantized in units of $e$. 
The value of $Q$ is proportional to the number of electron levels which pass through the Fermi level when the potential changes from $U_{min}$ to $U_{max}$. 

Turning to our dot model with an equidistant spectrum, the situation is even better.
For a periodic density of states, $\nu_{0}(E) = \nu_{0}(E+\Delta)$, we conclude from Eq.~(\ref{mb_4}) that the emitted charge is quantized, $Q=en$, if the difference $e\delta U = eU_{max} - eU_{min}$ is exactly equal to an integer number of level spacings, $e\delta U = n\Delta$.  
This holds for arbitrary QPC transparency $T$ and temperature $\theta$. 

As an illustration we calculate the charge $Q$ emitted for half of a period using Eq.~(\ref{mb_2}) for a monochromatic excitation $U(t) = U_{1}\cos(\Omega t)$ and an equidistant spectrum and plot the dependence $Q(U_{1})$ in Fig.~\ref{fig1}. 
In addition we plot the corresponding amount of charge carried by the first harmonic of the current. 
From Fig.~\ref{fig1} we can see that the first harmonic of the current reflects correctly the quantization of the emitted charge only within the first plateau. At higher excitation amplitudes the higher harmonics of the current need to be taken into account.

Importantly, the charge quantization is perfect at $e\delta U=n\Delta$. 
However if the potential difference deviates from this special value, $eu = e\delta U-n\Delta > 0$, then the deviation of the charge from the quantized value, $\delta Q = Q - en$, will depend on both the transmission $T$ and the temperature $k_B\theta$. 

For $\theta=0$ and $T\to 0$ 
in the special case that the Fermi level $\mu$ lies exactly in the middle between two levels of a dot at $U=U_{min}$ we find the plateau:  
$Q = e \left[\left[\frac{1}{2} + \frac{e\delta U}{\Delta} \right]\right]$,
where $[[X]]$ is the integer part of $X$.
At small temperatures, $k_B\theta\ll\Delta$ we have 
$\delta Q = 
2eu/(k_B\theta)\,\exp\{-\Delta/(2k_B\theta)\}$ at $eu \to 0$.
The violation of charge quantization is exponentially small at low temperatures unless we are at the transition point from one plateau to another, $\delta Q = \pm 1/2$ at  $eu=\Delta/2\mp 0$. 
Note to get $\delta Q$ at $eu \to \Delta$ one needs to replace $eu$ by $eu-\Delta$. 

Next we consider the effect of a non-zero QPC transmission probability on $\delta Q$ at zero temperature. 
At $T\ll 1$ the density of states for our model can be approximated as a sum of Breit-Wigner resonances. 
Then we find at $eu \to 0$:
$\delta Q = Teu/(\pi^2\Delta)$.
In contrast to the temperature, the effect of a finite QPC transmission is more crucial, since $\delta Q$ is linear in transmission $T$. 
Moreover, strictly speaking, the condition for the low frequency regime considered here is $\Omega\tau \ll T$. 
Therefore, with decreasing transmission we have to decrease the driving frequency to obtain charge quantization with a good accuracy.
To avoid the competition between the operating speed and the quantization accuracy one needs to tune the driving amplitude $e\delta U$ as close as possible to $n\Delta$.

{\it General current expression}.--
We will now derive a general expression for the time dependent current $I(t)$, valid for a pulse with arbitrary amplitude and frequency. 
For this we need the Floquet scattering matrix for a quantum dot subject to an arbitrary but periodic time dependent  $U(t)$.
It is useful to introduce the phase
$\Phi_{q}(t)=\frac{e}{\hbar}\int_{t-q\tau(E)}^{t}dt^{\prime}U(t^{\prime})$
accumulated in the time interval $t,t- q \tau$ it takes a carrier to execute $q$ round trips.
This carrier, exiting at time $t$ has entered the quantum dot at time $t-q \tau$. 
Thus we get,
\begin{equation}
\label{mm_2}
S_{in}(t,E) = r + \tilde{t}^2 \sum_{q=1}^{\infty}r^{q-1} 
e^{i\left\{q\varphi(E) - \Phi_{q}(t)\right\}}, 
\end{equation} 
where the phase
$\varphi(E) = \varphi(\mu) + \tau\hbar^{-1}(E-\mu)$ with $\tau=\tau(\mu)$.
Note that this is a Fabry-Perot type scattering matrix now including appropriate time-dependent phases. 
The Floquet scattering matrix is obtained from $S_{in}(t,E)$ through a Fourier transform 
$S_{F}(E_{n},E) = \int_{0}^{\cal T} \frac{dt}{\cal T}
e^{in\Omega t} S_{in}(t,E). $
In the stationary case, $U(t) = const$, only the term with $n=0$ remains. 
In this case we recover the stationary scattering amplitude 
$S_{F}(E,E) = S_{0}(E)$.

The current $I(t)$ in terms of $S_{F}$ is presented in Ref.~\cite{MB07}.
For the subsequent discussion it is convenient to write the current, $I(t)=I^{(l)}(t)+I^{(nl)}(t)$, as a sum of linear and non-linear terms in $U(t)$. 
Using Eq.(\ref{mm_2}) we get
\begin{equation}
\label{Eq_2}
\begin{array}{c}
I^{(l)}(t) = \frac{e^2}{h} T^2
\sum\limits_{q=1}^{\infty}R^{q-1}\left\{U(t) - U(t-q\tau) \right\} ,\\ 
\ \\
I^{(nl)}(t) = \frac{eT^2}{\pi\tau}\Im  \sum\limits_{p=1}^{\infty}\sum\limits_{q=1}^{\infty}\frac{1}{p}\eta\left(p\frac{\theta}{\theta^{\star}}\right)r^p e^{ip\varphi(\mu)}R^{q-1} \\
\ \\
\times\left(e^{-i\Phi_{p}(t-q\tau)} - e^{-i\Phi_{p}(t)}\right). 
\end{array}
\end{equation}
Here $R=|r|^2$ and $T=|\tilde t|^2$ are the reflection and transmission probabilities of the QPC,
$k_B\theta^{\star}=\Delta/(2\pi^2)$ and 
$\eta(x) = x/\sinh(x)$.
Taking the low frequency limit of Eq.~(\ref{Eq_2}) we arrive at our main result Eq.~(\ref{mb_2}).

{\it Square pulse excitation}.--
With Eq.~(\ref{Eq_2}) we can now investigate the real-time response to a square pulse excitation of duration $\cal T$ which is of much experimental interest. 
Suppose that at the time $t_0$ the potential changes from zero to $\delta U$. 
This change occurs on a short time scale $\delta t\ll \tau$.
Therefore, it is an essentially non-adiabatic excitation even if the pulse duration ${\cal T}\gg\tau$.   
Experimentally it was found for $e\delta U = n\Delta$ at low temperatures that the current decays exponentially in time until the next potential jump occurs~\cite{Feve07}.
The theory developed in Ref.~\cite{Feve07} for a square pulsed excitation gives a relaxation current $I(t) = q/\tilde\tau e^{-(t-t_0)/\tilde\tau}$ with an initial charge $q$ and a time constant $\tilde\tau$.
In the high temperature limit $\theta\gg\theta^{\star}$ or for the case $e \delta U = n \Delta$ these parameters are predicted to be $q = e^2\delta U/\Delta$ and $\tilde\tau = (2-T)h/(2T\Delta )$. 

The calculations based on our general expression, Eq.~(\ref{Eq_2}), 
show the following: 
At high temperatures $I^{(nl)}(t)$ vanishes and we find (${\cal T}>t-t_0>0$):
\begin{equation}
\label{Eq_11}
I(t) =  e^2\delta U T R^{N(t)}/h, \quad \theta\gg\theta^{\star},
\end{equation} 
where the integer $N(t) = \left[\left[(t-t_0)/\tau \right]\right]$.
The current decreases with time in a step-like manner being constant over a time interval $\tau$. 
At $t-t_0\gg \tau$ we can write $I(t) \sim e^{-(t-t_0)/\tau_D}$ with decay time $\tau_D = h/(\Delta\ln(1/R))$. 
This agrees well with Ref.~\cite{Feve07} since $\tau_{D}\approx \tilde\tau$ unless $T\sim 1$. 
Eq.~(\ref{Eq_11}) leads to an emitted charge: $Q = e^2\delta U/\Delta$, i.e. at high temperatures the emitted charge is not quantized. 

In contrast, at low temperatures, Eq.~(\ref{Eq_2}) leads to a quantized emitted charge. 
Its value depends on the potential change $\delta U$ in the same way as for a monochromatic excitation (see, Fig.~\ref{fig1}). 
The part $I^{(l)}$ of the total current remains the same for low and high temperatures. 
Therefore, the contribution due to $I^{(nl)}$ is crucial for the appearance of charge quantization at low temperatures. 

{\it Noise}.--
We next examine the noise of this fast electron emitter.
For a periodically driven quantum conductor the symmetrized correlation function 
${P}(\omega,\omega^{\prime}) = \sum_{l=-\infty}^{\infty} 2\pi \delta(\omega + \omega^{\prime} - l\Omega) {\cal P}_{l}(\omega)$
for currents at frequencies $\omega$ and $\omega^{\prime}$ depends on both frequencies. 
The expression for 
${\cal P}_{l}(\omega)$ in terms of the Floquet scattering matrix elements is presented in Ref.~\cite{MB07}.
Using Eq.~(\ref{mm_2}) 
we find for $\Omega\tau,\omega\tau\ll 1$ 
and at low temperatures, $k_B\theta\ll \hbar\Omega,\hbar\omega$:
\begin{equation}
\label{Eq_15}
{\cal P}_{l}(\omega) = \omega(\omega-l\Omega)
\frac{h^2e^2}{4\pi}\sum\limits_{q=-\infty}^{\infty}
\left|(l-q)\Omega -\omega \right| \nu_{l-q}\nu_{q} ,
\end{equation}
where
$\nu_{l-q}$ and $\nu_{q}$ are Fourier coefficients of the frozen density of states $\nu(t,E)$.
At higher temperatures or at high observation frequency
we can represent the noise in the form which is closely related to the noise of a capacitor in the linear response regime (see Ref~\cite{BTP93}):
\begin{equation}
\label{Eq_16}
\begin{array}{c}
{\cal P}_{l}(\omega) = \omega(\omega - l\Omega) 
\left\{C_{q}^{2} R_{q} \right\}_{l} 
\left\{
\begin{array}{ll}
\hbar|\omega|, &  \hbar\omega\gg\hbar\Omega, k_B\theta, \\
\ \\
2k_B\theta, &  k_B\theta\gg \hbar\Omega, \hbar\omega ,
\end{array}
 \right. \\
 \ \\
C_{q}^{2}(t) R_{q}(t) = \frac{he^2}{2} \int dE \left(-\frac{\partial f_0}{\partial E} \right)
\nu^2(t,E).
\end{array}
\end{equation}
Notice that the quantity $R_{q}(t)$ introduced above is different from $R_{\partial}(t)$ given in Eq.~(\ref{mb_3}). 
At zero temperature $R_q$ tends to its quantized linear response value $h /2e^{2}$. 

\begin{figure}[t]
  \vspace{0mm}
  \centerline{
   \epsfxsize 8cm
   \epsffile{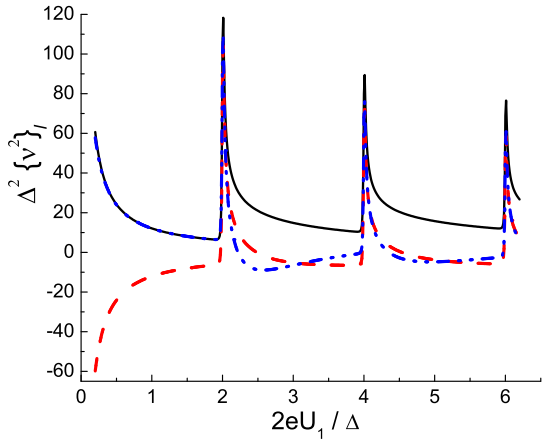}
             }
  \vspace{0mm}
  \nopagebreak
  \caption{ (Color online) The Fourier harmonics of the squared frozen density of states $\{\nu^2(t,\mu)\}_{l}$ in units of $1/\Delta^2$ calculated at the Fermi energy as a function of $U_1$ determine the noise of the emitter. $l = 0$ (solid black line); $2$ (dashed red line); $4$ (dash-dotted blue line).
For the parameters chosen (the same as in Fig.~\ref{fig1}) the odd harmonics are absent.}
\label{fig2}
\end{figure}

In Fig.~\ref{fig2} we plot several Fourier harmonics for the squared density of states [which defines the noise power Eq.~(\ref{Eq_16})] as a function of the amplitude of a monochromatic potential for the same parameters as in Fig.~\ref{fig1}. 
As one can expect the noise power peaks at those amplitudes which correspond to transitions from one plateau to another.
Correspondingly the noise power is minimal in the region where the emitted/absorbed charge is quantized, i.e., where the quantum dot emits electrons (holes) regularly.
For a regularly emitting source the zero-frequency noise power is zero. However the quantum capacitor under consideration does not produce a zero-frequency noise at all. 
Instead, as we see, the frequency dependent noise can be used to detect whether the particles are emitted regularly or not.  
  
In conclusion we have developed a non-linear dynamical scattering theory for a recently discovered high frequency quantized electron emitter.
Our theory allows to investigate analytically properties of such an emitter and to optimize them over a wide range of parameters.

We acknowledge discussions with Christian Glattli and Gvendal F\`{e}ve and thank Simon Nigg for a critical reading of the manuscript.
This work was supported by the Swiss NSF, the Swiss Center for excellence MaNEP and the Swedish VR.

\end{document}